\pdfoutput=1
\documentclass{svmult}
\usepackage{graphicx}
\usepackage{amsmath}
\usepackage{amssymb}
\usepackage{bm}
\usepackage[normalem]{ulem}

\newcommand{\ttt}[1]{``#1''}

\newcommand{\album}[1]{\uline{#1}}

\begin{document}

\title*{A Grateful Dead Analysis: The Relationship Between Concert and Listening Behavior}
\author{Marko A. Rodriguez$^1$ \and Vadas Gintautas$^1$ \and Alberto Pepe$^2$}

\institute{
	T-7, Center for Non-Linear Studies \\ 
	Los Alamos National Laboratory \\
	Los Alamos, New Mexico 87545 \\
	\texttt{marko@lanl.gov, vadasg@lanl.gov}\\
	\and
	Center for Embedded Networked Sensing \\ 
	University of California at Los Angeles \\
	Los Angeles, California 90095 \\
	\texttt{apepe@ucla.edu} 
}

\maketitle

\begin{abstract}
The Grateful Dead were an American band that was born out of the San Francisco, California psychedelic movement of the 1960s. The band played music together from 1965 to 1995 and is well known for concert performances containing extended improvisations and long and unique set lists. This article presents a comparative analysis between 1,590 of the Grateful Dead's concert set lists from 1972 to 1995 and 2,616,990 last.fm Grateful Dead listening events from August 2005 to October 2007. While there is a strong correlation between how songs were played in concert and how they are listened to by last.fm members, the outlying songs in this trend identify interesting aspects of the band and their fans 10 years after the band's dissolution.
\end{abstract}

\section{Introduction}

\begin{figure}[ht!]
\begin{center}
\includegraphics[width=0.175\textwidth]{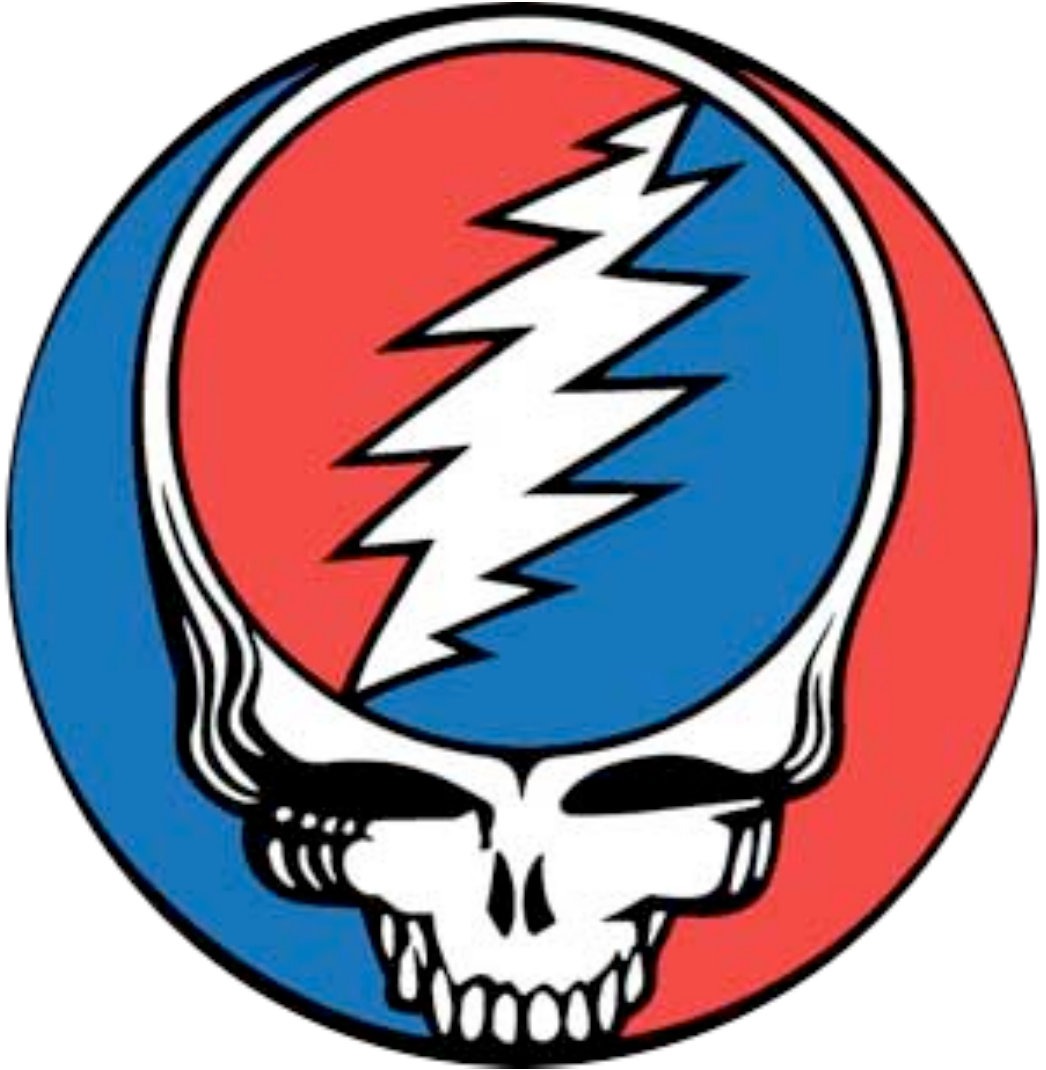}
\caption{\label{fig:stealyourface}The Grateful Dead ``Steal Your Face'' icon.}
\end{center}
\end{figure}

The Grateful Dead were an American band which, despite relatively little popular radio airtime, enjoyed a cult-like following from a fan base that numbered in the millions \cite{trip:mcnally2002}. The Grateful Dead originated in San Francisco, California in the early 1960s and toured the world playing concerts until the untimely death of the foreman and lead guitarist Jerry Garcia in 1995. The primary source of revenue and exposure for the band came through their concert tours.   They played over 37,000 songs live, in some 2,300 concerts over their 30 years as a band~\cite{setlists:1996}. Throughout their years together, the Grateful Dead accumulated a large repertoire that included over 450 unique songs \cite{setlists:1996}. The Grateful Dead's success and continuity across multiple generations of music listeners is perhaps due in part to their fundamentally eclectic nature. The band utilized many song writers, composers and singers, and this resulted in a broad diversity in sound. Robert Hunter and John Barlow were the primary lyricists for scores written by Jerry Garcia and Bob Weir, respectively \cite{annotatedead:dobb2007}. While Jerry Garcia and Bob Weir were the primary singers as well, other singers included Ron McKernan, Brent Mydland, and Phil Lesh. Moreover, their eclectic nature can be seen in the large number of graphic icons they used to represent themselves. These icons include skeletons, roses, dancing bears, terrapins, etc. Perhaps their most famous and recognizable image is the ``Steal Your Face'' icon in Figure \ref{fig:stealyourface} that was released as the album cover art to the live 1976 \album{Steal Your Face} album \cite{stealface:dead1976}.

The history of the Grateful Dead's album releases (13 studio albums and 77 live albums) further reinforces the band's emphasis on concerts.  More live albums are released regularly as high quality recordings of good performances are discovered in the Grateful Dead concert archive. For the band and for the fans, the performances of the Grateful Dead were all about diversity in the live music experience. In any given show, the concert set list, the improvisations, and the mood of the band all varied. In concert, all of these factors came together to create a unique experience for their fans each and every time. 

Perhaps even more astounding than their prolific concert performances is the dedication that their fans (known as ``deadheads'') had to their music \cite{deadheads:grushkin1983,adams:deadhead1998,sardiello:deadhead, deadhead:pattacini}. The typical deadhead was not a passive consumer of recorded studio albums, but an active concert goer that traveled with the band from concert to concert, city to city, and country to country. Some 10 years after the Grateful Dead disbanded, the band's music is still heavily listened to as evinced by statistics gathered from the popular online music service known as last.fm.\footnote{last.fm is available at: \texttt{http://www.last.fm/}} The last.fm ``audioscrobbler'' plug-in is recommendation software that works with popular computer music players such as iTunes or Winamp.  Whenever a song is played using, say, iTunes, the plug-in reports this activity to the last.fm server where it is aggregated. From August 2005 to October 2007, there were over 2.5 million Grateful Dead song usages recorded by last.fm. With 72\% of the users of last.fm under the age of 35\footnote{Source: last.fm internal web statistics, courtesy Anil Bawa-Cavia.}, the popularity of the Grateful Dead, a generation of fans later and 10 years after the band's dissolution, is still very strong. 

This article presents an analysis of the Grateful Dead's concert behavior and exposes a relationship between the concert song patterns from 1972 to 1995 and the last.fm listening statistics of the band's songs from August 2005 to October 2007. First the available set list data is summarized and presented with an analysis of the concert behavior of the band. Next the usage data from last.fm is presented with an analysis of the listening behavior of last.fm members. Finally, a comparative analysis of the concert and listening behavior of the Grateful Dead is presented.

\section{The Grateful Dead Concert Behavior\label{sec:concert}}

Concert set lists provide the raw data from which to study the concert behavior of the Grateful Dead.\footnote{Set list data obtained from \texttt{http://www.cs.cmu.edu/People/gdead/setlists.html}. The data were cleaned to remedy various typographical alterations (e.g.~\ttt{Trucking} and \ttt{{Truckin'}} are the same song), to fix various spelling errors (e.g.~\ttt{Warf Rat} and \ttt{Wharf Rat} are the same song), and to fix various abbreviations (e.g. \ttt{China Cat} and \ttt{China Cat Sunflower} are the same song).} The data gathered include 1,590 set lists for concerts from 1972 to 1995, with 28,904 individual song plays. A typical, unmodified set list is presented below:
\begin{footnotesize}
\begin{verbatim}
Winterland Arena, San Francisco, CA (12/31/77)

Music Never Stopped
Tennessee Jed
Funiculi Funicula
Me and My Uncle
Loser
Jack Straw
Friend of the Devil
Lazy Lightnin'
Supplication

Sugar Magnolia
Scarlet Begonias
Fire on the Mountain
Truckin'
Wharf Rat
drums
Not Fade Away
Around and Around

One More Saturday Night
Casey Jones
\end{verbatim}
\end{footnotesize}
Blank lines divide the set list into 4 components. The first component is the concert venue and location along with the date that the concert was played. The second component is the first set list in the sequence in which the songs were played. For example, \ttt{Friend of the Devil} was played immediately after \ttt{Jack Straw} in the above example. The third component is the second set list of the concert. The second set of the Grateful Dead is known for fewer songs and extended improvisational sessions. Furthermore, the second set of a Grateful Dead concert is known for its ``blending'' of songs in which there exist fewer pauses between the end of one song and the beginning of the next; that is, the second set was often a large medley of sorts.  Second set medleys often featured pairs of songs that were almost always played together.  For example, \ttt{China Cat Sunflower} almost always preceded \ttt{I Know You Rider}, but never in the opposite order.  Also, \ttt{China Cat Sunflower} was very rarely played with a different song following.\footnote{A similar pattern exists for other song pairs such as \ttt{Scarlet Begonias} and \ttt{Fire on the Mountain}, \ttt{Saint of Circumstance} and \ttt{Lost Sailor}, and \ttt{Cryptic Envelopment} and \ttt{The Other One}.}  The fourth and last component, which is usually the shortest, is the encore set list. The Grateful Dead were known to typically play their concerts in this 3 set form.  

\begin{figure}[h!]
    \begin{minipage}{2.75in}
\begin{center}
\begin{footnotesize}
\begin{tabular}{lcll|}
\hline
\textbf{song}&\textbf{times played}\\\hline
\ttt{Drums}&1386\\
\ttt{Playing in the Band}&651\\ 
\ttt{Sugar Magnolia}&494\\ 
\ttt{Not Fade Away}&486\\ 
\ttt{The Other One}&438\\ 
\ttt{Jack Straw}&437\\ 
\ttt{Trucking}&427\\ 
\ttt{Me and My Uncle}&412\\ 
\ttt{Looks Like Rain}&407\\ 
\ttt{Promised Land}&407\\ 
\ttt{I Know You Rider}&406\\ 
\ttt{China Cat Sunflower}&403\\ 
\ttt{New Minglewood Blues}&398\\ 
\ttt{Around and Around}&395\\ 
\ttt{Tennessee Jed}&390\\
\hline
\end{tabular}
\caption{\label{tab:plays}The top 15 Grateful Dead songs played in concert from 1972 to 1995.}
\end{footnotesize}
\end{center}
\end{minipage}
\qquad
\begin{minipage}{2.75in}
\begin{center}
\includegraphics[width=1\textwidth]{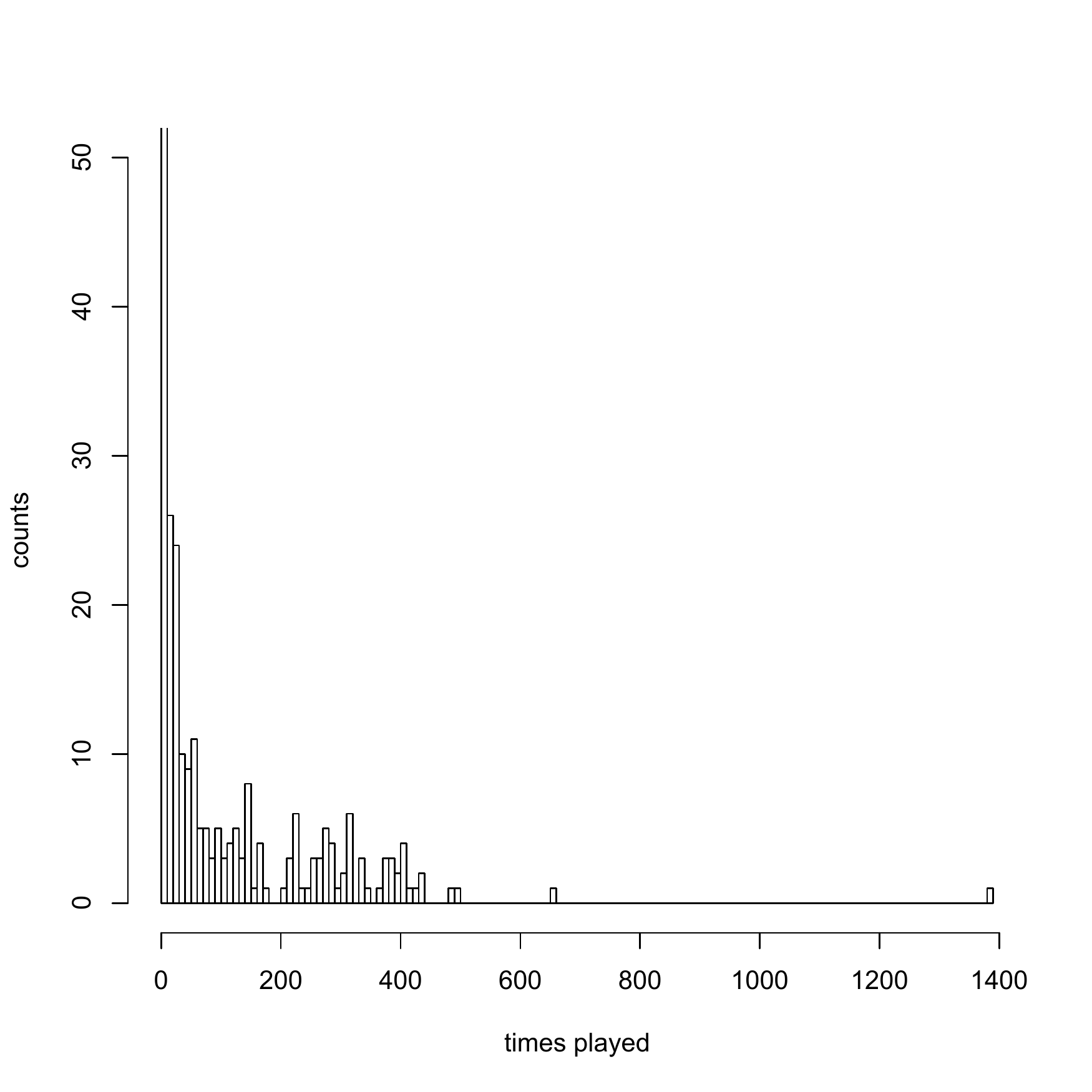}
\caption{\label{fig:plays-distribution}A histogram of the number of times a song was played in concert. Most songs were played only a few times and very few songs were played many times. (The vertical axis was trimmed from its maximum of 150 in order to preserve diagram clarity).}
\end{center}
\end{minipage}
\end{figure}

A basic measure, calculated using many concert set lists, is to simply count the number of times a given song is played over all concerts.  This ranked list of songs is a rudimentary ``greatest hits'' list of sorts, but also a histogram of concert plays sheds light on the distribution of these counts.  Did most songs get approximately the same number of concert plays, or did the band play a small set of favorite songs interspersed with less popular songs to provide variety? Table~\ref{tab:plays} shows the raw counts for the 15 most played songs.  Note that of the 1,590 concerts analyzed, 1,386 of those concerts included the \ttt{Drums} improvisational rhythm sequence, which typically appeared in the second set of most concerts.  The Grateful Dead very often used \ttt{Drums} in the second set to bridge two songs that would not otherwise be simple to link as a medley.  Figure \ref{fig:plays-distribution} presents a histogram denoting the number of songs that were played a given number of times. In summary, many songs were played only a few times and few songs were played many times.

\section{The Grateful Dead Usage Statistics\label{sec:usage}}

\begin{figure}
    \begin{minipage}{2.75in}%
\begin{center}
\begin{footnotesize}
\begin{tabular}{lcll|}
\hline
\textbf{song}&\textbf{times used}\\\hline
\ttt{Friend of the Devil}&143988\\
\ttt{Sugar Magnolia}&124736\\
\ttt{Trucking}&122877\\
\ttt{Casey Jones}&102449\\
\ttt{Box of Rain}&88340\\
\ttt{Uncle John's Band}&82431\\
\ttt{Ripple}&80629\\
\ttt{Touch of Grey}&71270\\
\ttt{Brokedown Palace}&54675\\
\ttt{Candyman}&54344\\
\ttt{Fire on the Mountain}&48516\\
\ttt{Franklin's Tower}&45404\\
\ttt{Scarlet Begonias}&42137\\
\ttt{Dark Star}&39953\\
\ttt{China Cat Sunflower}&36479\\
\hline
\end{tabular}
\caption{\label{tab:usage}The top $15$ downloads of Grateful Dead songs on last.fm from August 2005 to October 2007.}
\end{footnotesize}
\end{center}
\end{minipage}
\qquad
\begin{minipage}{2.75in}%
\begin{center}
\includegraphics[width=1\textwidth]{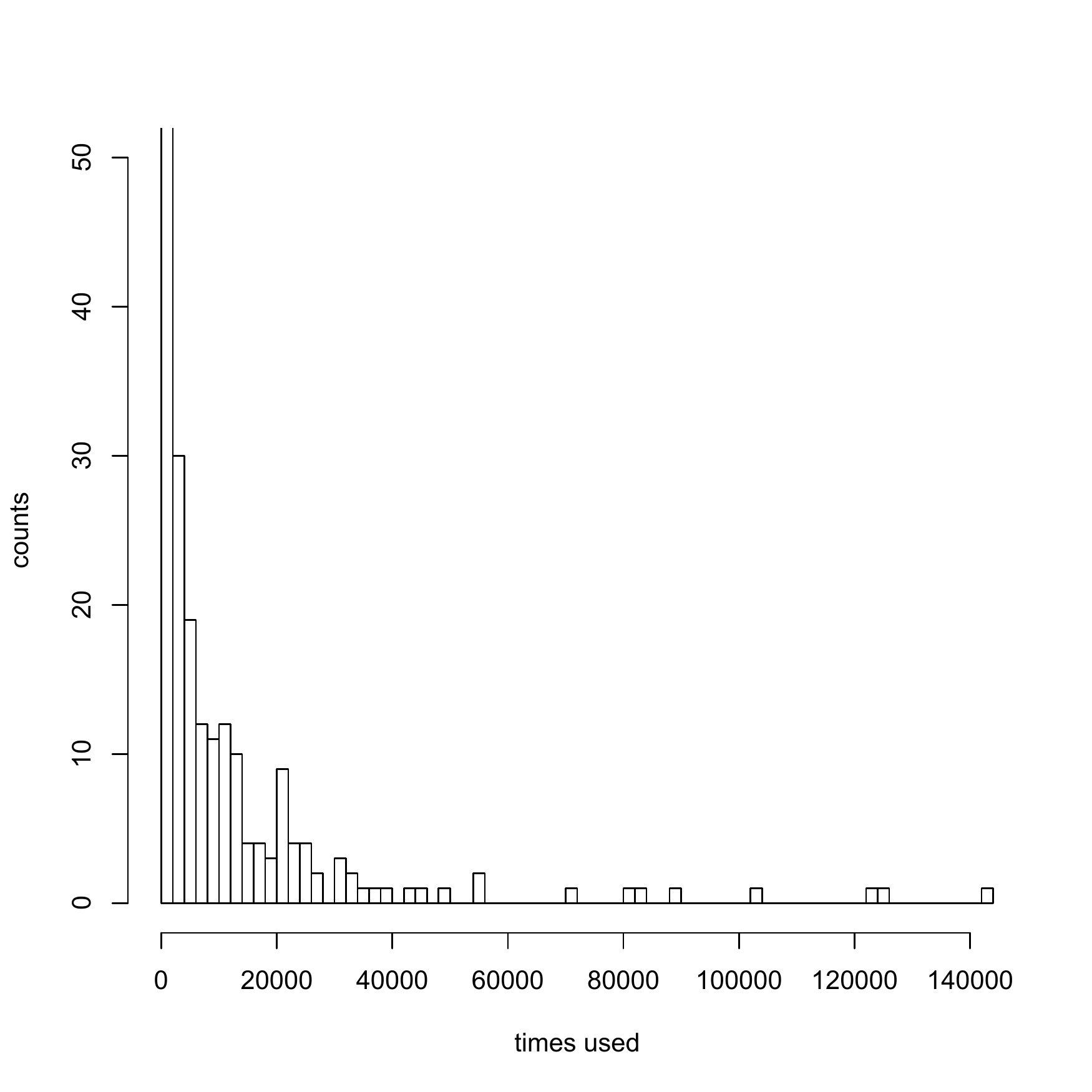}
\caption{\label{fig:usage-distribution}A histogram of times a song was listened to, from last.fm data. (The vertical axis was trimmed from its maximum of 200 in order to preserve diagram clarity.)}
\end{center}
\end{minipage}
\end{figure}

The online music service last.fm tracks how registered members enjoy music by what songs they play on Internet radio or through a computer music player plug-in.  The last.fm service maintains a database of the listening behavior of its registered members. From this database, the last.fm service is able to recommend songs and artists to its members based on the listening behavior of similar members.  This service is analogous to Amazon.com using historical purchasing behavior to recommend products to customers. Table \ref{tab:usage} lists the top $15$ Grateful Dead songs listened to by last.fm members. These data were gathered from August 2005 to October 2007 and include 2,616,990 unique listening events.  Figure \ref{fig:usage-distribution} presents a histogram of the listening counts of songs. In summary, similar to the concert behavior of the Grateful Dead, many songs were listened to a few times and a few songs were listened to many times.

\section{The Relationship Between Concert and Usage Behavior\label{sec:both}}

The last.fm music service records listening behavior of songs that have appeared in some published form, such as studio albums, singles, and recordings of live concerts. The Grateful Dead deviated from the standard model of releasing studio albums, in that their primary revenue stream was through concert performances, even from the start. The Grateful Dead produced 13 studio albums and 77 live albums.  The band's first live album, \album{Live/Dead}, was released in 1969 \cite{livedead:dead1969}.  With respect to the influence of live recordings on the present generation of listeners, nearly all live albums were a direct reflection of a particular live concert performance and as such, respected that concert set list's song sequence. Thus, if a last.fm listener were to listen to any one of the many live albums, he or she is, in fact, replaying concert history and contributing proportionately to the number of songs listened to as times played in concert. This notion is further accentuated by fans that created digital renditions of their favorite concert tapes.\footnote{Prior to the advent of the Internet and the easy distribution of digital audio files, Grateful Dead concert tape trading was an extremely popular way of disseminating the Grateful Dead's live experience. In fact, the band encouraged this \cite{tapers:dwork1998}. Many tapes have now been digitized and shared online by fans.} Given that there are currently 5 times more live albums than studio albums, one may expect that last.fm users would primarily listen to recordings of concerts and that the usage data would be directly correlated with that of the set list data.  However, as shown in this section, there are significant deviations from a perfect correlation.  Reasons for these deviations may include the fact that listeners can replay only their favorite songs from live albums, may prefer studio albums to recordings of live shows, and are able to make compilations of tracks (``playlists'') that differ from the live and studio productions. 

\begin{figure}[ht]
\begin{center}
\includegraphics[angle=90,width=1.0\textwidth]{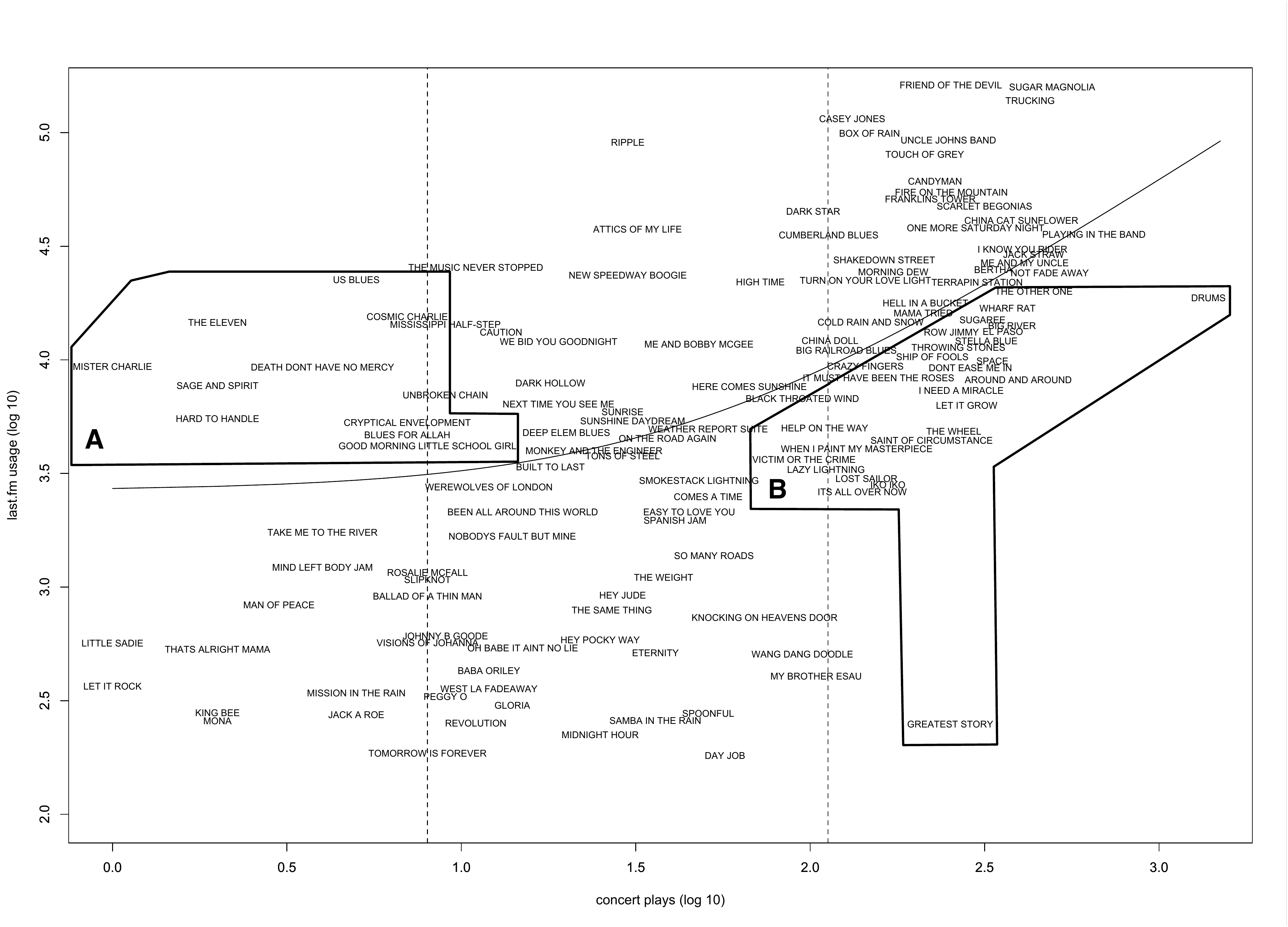}
\caption{\label{fig:usage-plays}Grateful Dead concert plays vs. last.fm usage. Unfortunately, not all song names could be displayed due to severe overlapping of the labels. In order to remedy this situation, in the more dense regions, song labels were randomly removed.}
\end{center}
\end{figure}

Figure \ref{fig:usage-plays} plots each song in a two-dimensional space. Each song is provided a coordinate in this space, where the horizontal coordinate is the number of times the song was played in concert and the vertical coordinate is the number of times the song was listened to by last.fm members.
Given that each song has two associated values (horizontal and vertical coordinates), it is possible to form two vectors of these numbers and to measure the correlation of these vectors to determine how strongly their values are related. The analysis reveals that the listening behavior of last.fm members is strongly correlated with the concert behavior of the Grateful Dead (a measured correlation of $0.763$ where $0$ is uncorrelated and $1$ is perfectly correlated).\footnote{A Spearman $\rho$ rank-order correlation reveals a correlation of $\rho = 0.763$ with a $p$-value $< 2.2\times 10^{-16}$, where $\rho = -1$ is inversely related, $\rho=0$ is unrelated, and $\rho=1$ is correlated.  The $p$-value indicates the probability of such a correlation occurring randomly; the $p$ value for this correlation indicates that, at random, this correlation would occur 5 times out of $10^{15}$~\cite{parame:sheskin2004}.}  Since the data are not perfectly correlated, it is valuable to examine songs which are exceptions to this trend.

The solid curved line in Figure \ref{fig:usage-plays} shows regions for which songs display a strong correlation.\footnote{Specifically, the line represents a linear regression model that predicts usage in terms of concert plays. The data were fit to a line; the line appears curved because the values on both axes are plotted on logarithmic (base-$10$) scale to preserve diagram clarity.} The two dashed lines represent the $40^{\text{th}}$-quantile and $75^{\text{th}}$-quantile of concert plays. These quantiles were subjectively chosen in order to provide a discussion of songs that are noteworthy outliers from the trend. Finally, the two closed polygons labeled {\bf A} and {\bf B} represent the interesting outlying songs. Polygon {\bf A} encapsulates songs that were heavily listened to by last.fm members, but played few times in concert by the Grateful Dead. On the other hand, {\bf B} encapsulates songs that were heavily played by the Grateful Dead, but less listened to (and possibly overlooked) by last.fm members.\footnote{A random sample of songs were removed from this diagram to preserve clarity as to ensure that song labels did not overlap. Furthermore, while {\bf A} and {\bf B} occupy regions outside of their quantile delimitations, this is to ensure that song labels were encapsulated.}

Polygon {\bf A} encapsulates those songs that were heavily listened to by last.fm users but not heavily played by the Grateful Dead in concert. Many of the songs in polygon {\bf A} are old classics that did not persist due to either changes in the band or to the band's extremely prolific period of songwriting in the early 1970s. For example, although the song \ttt{Mister Charlie} was played for the last time in concert on May 26, 1972 at the Strand Lyceum in London, England, it was featured on the \album{Europe `72} album \cite{europe:dead1972}.  This album reached a peak spot of 24 on the Billboard pop albums chart in 1973.  Thus, while not being played much in concert during the band's lifetime, \ttt{Mister Charlie} remains a fan treasure by virtue of securing a place on a much celebrated album. Similar arguments can be made for \ttt{Sage and Spirit} and \ttt{Blues for Allah} which, while technically difficult and therefore generally avoided in concert, are songs that were released on the \album{Blues For Allah} LP \cite{allah:dead1975} which reached spot 12 on the Billboard charts for pop album in 1975. \ttt{Good Morning Little School Girl} was primarily sung by Ron McKernan, whose untimely death in 1973 caused the song to be removed from the Grateful Dead concert play list except for a few special appearances in the late 80s and early 90s.\footnote{It is worth noting that \ttt{Mister Charlie} was also primarily sung by Ron McKernan and thus, didn't last with the Grateful Dead past his lifetime.} However, \ttt{Good Morning Little School Girl} is the opening track of the \album{Two from the Vault} album \cite{twovalut:dead1992} that contains a live recording of the August 24, 1968 Shrine Exhibit Hall performance of the Grateful Dead. Although Jerry Garcia was the primary singer of \ttt{Death Don't Have No Mercy}, the song had a similar concert fate as \ttt{Good Morning Little School Girl} and was also released on the \album{Two from the Vault} live album.  \album{Two from the Vault} reached position 119 in 1992 on the Billboard charts.\footnote{\ttt{The Eleven} and \ttt{Cryptic Envelopment} are also on \album{Two from the Vault}.} Likewise, \ttt{Hard to Handle}, like \ttt{Good Morning Little School Girl} was a Ron McKernan specialty that was dropped after his death, except for being played during the New Year's Eve show of 1982 in Oakland, California. Finally, \ttt{Cryptic Envelopment} provided a medley prelude to the popular \ttt{The Other One} and only later in the bands life was \ttt{The Other One} separated from \ttt{Cryptic Envelopment} and preceded by \ttt{Drums}. It is worth noting that 28 of the 77 live albums of the Grateful Dead include \ttt{The Other One} while only 7 include \ttt{Cryptic Envelopment}. 

Polygon {\bf B} encapsulates those songs that were heavily played in concert throughout the Grateful Dead's career, but for various reasons, were less frequently listened to by last.fm members. \ttt{Drums} is perhaps the most salient of this collection of songs at the extreme of the boundary. \ttt{Drums} is an all-drum improvisational piece that usually appeared in the second set of a Grateful Dead concert. Usually appearing with \ttt{Drums} was the full sonic spectrum improvisation of \ttt{Space} which included all band members. Both \ttt{Drums} and \ttt{Space} found a stable home on the \album{Infrared Roses} album \cite{infrared:dead1991}, but unfortunately, due to the esoteric nature of these improvisations, Infrared Roses has been less well received by the general public and thus received no popular awards and did not make it on any music charts. Furthermore, to compound the situation, the Grateful Dead provided unique names for \album{Infrared Roses} tracks and thus, when played by last.fm users, are not associated with the typical \ttt{Drums} and \ttt{Space} songs of the concert set lists. It is interesting to note the songs \ttt{Saint of Circumstance,} \ttt{When I Paint My Masterpiece,} \ttt{Victim or the Crime,} \ttt{Lost Sailor,} and \ttt{Greatest Story} in the bottom left of polygon {\bf B}. All of these songs were created by the song writing duo of Barlow and Weir and sung in concert often by Bob Weir. While these songs were played extensively in concert, they received relatively little attention from last.fm users. 

Finally, the extreme upper right of this plot is important as \ttt{Trucking} and \ttt{Sugar Magnolia} represent not only the most popular songs in terms of times played in concert, but in terms of times listened to on last.fm. \ttt{Trucking} is on 25 of the 90 released Grateful Dead albums and \ttt{Sugar Magnolia} is on 32 of those albums. Both \ttt{Trucking} and \ttt{Sugar Magnolia} were also well received publicly. \ttt{Trucking} reached position 64 in 1971 and \ttt{Sugar Magnolia} reached position 91 in 1973 on the Billboard pop singles charts. Also in this area is \ttt{Touch of Grey}. \ttt{Touch of Grey} was the only Grateful Dead song with an accompanying music video and in 1987, reached the top 10 Billboard single's chart. By comparison to produced greatest hits albums, Table \ref{tab:hits1} lists the songs that were released on the 2003 \album{Very Best of the Grateful Dead} compilation. Of these songs, 12 out of the 17 songs are in the top right quadrant (these songs are marked with an $^*$ in Table \ref{tab:hits1}), meaning that they were both played and listened to heavily by the Grateful Dead and their fans, respectively.\footnote{\ttt{Eyes of the World} and \ttt{Estimated Prophet} are not displayed as they were randomly removed to preserve diagram clarity.} A similar situation exists with the \album{Skeletons from the Closet} greatest hits album for which 8 out of 11 songs are in the top right quadrant \cite{skelcloset:dead1974} (Table \ref{tab:hits2} presents the songs on the album).\footnote{\ttt{Mexicali Blues} is not displayed as it was randomly removed to preserve diagram clarity.} Of particular importance is \ttt{Box of Rain} (on \album{Very Best of the Grateful Dead}) by Phil Lesh and Robert Hunter. This song, written after the death of Phil Lesh's father, is not only the last song ever played by the Grateful Dead in concert\footnote{The final Grateful Dead performance took place at Soldier Field on July 9, 1995 in Chicago. Jerry Garcia died exactly one month later on August 9, 1995.}, but also unique in that it is one of the few songs for which Phil Lesh was the primary singer. In summary, the upper right hand quadrant of this diagram is ripe for creating compilation and greatest hits albums as it reflects both what the band as well as present day fans appreciate.
\begin{table}
    \begin{minipage}{2.75in}
    \begin{center}
\begin{footnotesize}
    \begin{tabular}{lll}
\hline
\textbf{\#}&\textbf{\hspace{0.1in}Track name}\\\hline
1&\ttt{Trucking} $^*$\\ 	 	
2&\ttt{Touch of Grey} $^*$\\
3&\ttt{Sugar Magnolia} $^*$\\
4&\ttt{Casey Jones} $^*$\\ 	
5&\ttt{Uncle John's Band} $^*$\\
6&\ttt{Friend of the Devil} $^*$\\
7&\ttt{Franklin's Tower} $^*$\\
8&\ttt{Estimated Prophet} $^*$\\	
9&\ttt{Eyes of the World} $^*$\\	
10&\ttt{Box of Rain} $^*$\\	
11&\ttt{U.S.  Blues}\\
12&\ttt{The Golden Road to Ultimate Devotion}\\
13&\ttt{One More Saturday Night} $^*$\\
14&\ttt{Fire on the Mountain} $^*$\\
15&\ttt{The Music Never Stopped}\\
16&\ttt{Hell in a Bucket}\\
17&\ttt{Ripple}\\
\hline
\end{tabular}
\caption{\label{tab:hits1}The tracks of the \album{Very Best of the Grateful Dead} greatest hits album \cite{hits:dead2003}.  Upper right quadrant songs in Figure~\ref{fig:usage-plays} are marked with $^{*}$.}
\end{footnotesize}
\end{center}
    \end{minipage}
    \qquad
    \begin{minipage}{2.75in}
    \begin{center}
\begin{footnotesize}
    \begin{tabular}{lll}
\hline
\textbf{\#}&\textbf{\hspace{0.1in}Track name}\\\hline
1&\ttt{The Golden Road to Ultimate Devotion}\\ 	
2&\ttt{Trucking} $^*$\\
3&\ttt{Rosemary}\\
4&\ttt{Sugar Magnolia} $^*$\\ 	
5&\ttt{St. Stephen}\\
6&\ttt{Uncle John's Band} $^*$\\
7&\ttt{Casey Jones} $^*$\\
8&\ttt{Mexicali Blues} $^*$\\	
9&\ttt{Turn on Your Love Light} $^*$\\	
10&\ttt{One More Saturday Night} $^*$\\	
11&\ttt{Friend of the Devil} $^*$\\
\\ \\ \\ \\ \\ \\
\hline
\end{tabular}
\caption{\label{tab:hits2}The tracks of the \album{Skeletons in the Closet} greatest hits album \cite{skelcloset:dead1974}. Upper right quadrant songs in Figure~\ref{fig:usage-plays} are marked with $^{*}$.}
\end{footnotesize}
\end{center}
    \end{minipage}
\end{table}

\section{Conclusion}

The Grateful Dead were an American music phenomenon that influenced multiple generations of music lovers. For 30 years, the Grateful Dead made a career out of an unrelenting tour schedule that took them around the world, and, unlike typical bands, took their fans with them. It is now 43 years since the band started and with the use of online music providers and services, it is possible to track the listening behavior of the many Grateful Dead fans in the world today. This article presented an analysis comparing the popularity of Grateful Dead songs as identified by both how many times they were played in concert and how many times they were listened to by members of the last.fm online music service. The correlation between concert plays and fan listens is strong, but not perfect. Those songs that existed as outliers to a perfect correlation were analyzed to understand what made these songs deviate from the model. These deviations can be understood by changes in the band, by live performance album releases, and by the very nature of the songs themselves.

There is much to be learned about American concert tour culture and the bands that bring this culture to fruition \cite{concert:black2007}. Perhaps more than any other band, there exists large amounts of Grateful Dead data that go beyond set lists to include lyrics, chord progressions, concert reviews, and history. Many books have been published about the Grateful Dead and albums continue to be released 13 years after their final concert in 1995.  Without a doubt, the Grateful Dead have made a profound impact on that which is American rock music.

\section*{Acknowledgements}

The authors would like to thank Anil Bawa-Cavia of last.fm for providing the Grateful Dead raw usage data, Jerry Stratton for providing the set list data, and the Grateful Dead for doing what they do best.

\end{document}